\documentclass[aps,reprint,groupedaddress,nofootinbib]{revtex4-2}

\usepackage[T1]{fontenc} % if needed
\usepackage{subcaption}
\usepackage{verbatim}
\usepackage{mathrsfs}
\usepackage{physics}
\usepackage{amsthm}
\usepackage{mathtools}
\usepackage{hyperref}
\usepackage{enumerate}
\usepackage{amssymb}

\renewcommand{\tilde}{\widetilde}

\newcommand{\reals}{\mathbb{R}}
\newcommand{\comps}{\mathbb{C}}

\newcommand{\A}{\mathcal{A}}

\renewcommand{\H}{\mathcal{H}}

\newcommand{\B}{\mathcal{B}}

\renewcommand{\hat}{\widehat}

\usepackage[normalem]{ulem}
\usepackage{xcolor}

\begin{document}
	\title{Relative state-counting for semiclassical black holes}
	\author{Chris Akers}
	\affiliation{Institute for Advanced Study, 1 Einstein Dr, Princeton, NJ 08540}
	\author{Jonathan Sorce}
	\affiliation{Center for Theoretical Physics, Massachusetts Institute of Technology, 182 Memorial Drive, Cambridge, MA, USA}
	\date{\today}
	\begin{abstract}
		It has been shown that entropy differences between certain states of perturbative quantum gravity can be computed without specifying an ultraviolet completion.
		This is analogous to the situation in classical statistical mechanics, where entropy differences are defined but absolute entropy is not.
		Unlike in classical statistical mechanics, however, the entropy differences computed in perturbative quantum gravity do not have a clear physical interpretation.
		Here we construct a family of perturbative black hole states for which the entropy difference can be interpreted as a relative counting of states.
		Conceptually, this paper begins with the algebra of mass fluctuations around a fixed black hole background, and points out that while this is a type I algebra, it is not a factor and therefore has no canonical definition of entropy.
		As in previous work, coupling the mass fluctuations to quantum matter embeds the mass algebra within a type II factor, in which entropy differences (but not absolute entropies) are well defined.
		It is then shown that for microcanonical wavefunctions of mass fluctuation, the type II entropy difference equals the logarithm of the dimension of the extra Hilbert space that is needed to map one microcanonical window to another using gauge-invariant unitaries.
		The paper closes with comments on type II entropy difference in a more general class of states, where the von Neumann entropy difference does not have a physical interpretation, but ``one-shot'' entropy differences do.
	\end{abstract}
	
	\maketitle
	
	\section{Introduction}
	
	Entropy was first introduced in \cite{Clausius:entropy} as an empirical measure of change in thermodynamic processes.
	In this setting, it could be said that a system had gained or lost a certain amount of entropy, but absolute entropy was not meaningful.
	With classical statistical mechanics, Boltzmann \cite{Boltzmann} and Gibbs \cite{Gibbs} gave a physical interpretation of entropy exchange in terms of probability distributions of statistical systems.
	Due to an ambiguity in the choice of units on phase space, the Boltzmann-Gibbs theory gave well defined answers for entropy difference, but absolute entropies remained out of reach.
	This lacuna was eventually filled by quantum mechanics thanks to von Neumann's observation that for quantum density matrices, entropy is unambiguous \cite{Neumann}.
	
	Analogously, in general relativity, it was observed in the 1970s that perturbations of stationary black holes obey a law similar to the first law of thermodynamics, with the black hole's surface gravity playing the role of temperature, and its horizon area playing the role of entropy \cite{Bekenstein:1973ur, Bardeen:1973gs}.
	The constant of proportionality between these quantities was fixed by Hawking's calculation of a black hole's temperature \cite{hawking1975particle, Wald:thermal}, leading to the formula
	\begin{equation} \label{eq:first-law}
		\Delta S_{\text{BH}}
		= \frac{\Delta A}{4 G_N}
	\end{equation}
	in units with $k_B = \hbar = c = 1.$
	In analogy with quantum statistical mechanics, it was guessed that the absolute quantity $A/4G_N$ should represent the number of quantum gravity microstates making up an ensemble state represented by the classical black hole.
	This prediction has been verified for stationary black holes in many concrete models of quantum gravity, including in certain string theories \cite{Strominger:strings, Callan:micro, Horowitz:micro, Maldacena:micro, Johnson:micro}, in the AdS$_3$/CFT$_2$ correspondence \cite{Strominger:CFT, Hartman:light-spectrum}, and in Euclidean path integral approaches to holography \cite{Gibbons:1976ue, Penington:replica, Balasubramanian:2022gmo}.
	
	So far we have described a passage from black hole thermodynamics (i.e., classical black hole perturbation theory) to black hole quantum statistical mechanics (i.e., black holes in quantum gravity).
	In the context of ordinary thermodynamics, however, there was an intermediate theory --- classical statistical mechanics --- that could be used to compute and interpret differences in entropy without requiring absolute entropies to be defined.
	This prompts us to ask if there a theory that sits between classical black hole perturbation theory and microscopic quantum gravity, and that lets us compute and interpret entropy differences without needing to know how quantum gravity works.
	
	The theory that naturally sits between classical general relativity and microscopic quantum gravity is the theory of \textit{semiclassical} quantum gravity, in which general relativity is perturbatively coupled to quantum fields and in which metric fluctuations are quantized using the ordinary rules of quantum theory.
	Recently it has been shown that in certain settings within semiclassical quantum gravity, a mathematical framework known as Tomita-Takesaki theory provides an unambiguous definition of entropy difference \cite{Chandrasekaran:dS, Chandrasekaran:large-N, Penington:JT, Jensen:2023yxy, AliAhmad:2023etg, Kudler-Flam:2023qfl}.
	This entropy difference is associated with an abstract mathematical structure called a type II algebra.
	The most tantalizing aspect of this recent work is that when the type II entropy-difference formula is regulated, it can be interpreted as a difference between the horizon areas of two black holes, together with the difference between matter entropies of quantum fields in the black hole exteriors.
	This is a compelling result, but a key question remains: can the type II entropy difference be given an interpretation in terms of a relative counting of states?
	Because every nonzero operator in a type II algebra has infinite rank, it may seem difficult to formulate the idea that one density matrix is supported on more states than another.
	However, the analogy to classical statistical mechanics provides some hope that such an interpretation is possible, since in classical statistical mechanics entropy difference can be interpreted as the increased number of states consistent with one macroscopic configuration relative to another, even though the number of classical states consistent with any macroscopic configuration is formally infinite.
	
	The purpose of this paper is to explain a setting in which type II entropy difference admits a physical interpretation as a relative counting of semiclassical states.
	The main result is that for a Hilbert space describing mass fluctuations of a static black hole coupled to quantum fields, the type II entropy difference between wavefunctions describing microcanonical windows of mass fluctuation has the following interpretation.
	If $\H_{K_2}$ and $\H_{K_1}$ are the subspaces of the mass Hilbert space on which two such wavefunctions are supported, $\Delta S$ is the (assumed nonnegative) difference in their type II entropies, and $\H_{\text{QFT}}$ is the full Hilbert space of quantum fields, then for any integer $n \geq e^{\Delta S}$, the microcanonical semiclassical Hilbert space $\H_{\text{QFT}} \otimes \H_{K_2}$ can be embedded in the Hilbert space $\H_{\text{QFT}} \otimes \H_{K_1} \otimes \comps^n$ using a unitary embedding map that is constructed from operators in the type II algebra.
	For $n < e^{\Delta S},$ no such embedding is possible.
	
	Section \ref{sec:no-canonical} describes a framework for quantizing black hole mass in perturbative quantum gravity, and explains that there is no canonical formula for entropy or entropy difference.
	Section \ref{sec:crossed-product} gives a brief review of the work on gravity and the crossed product that was developed in \cite{Chandrasekaran:dS, Chandrasekaran:large-N, Penington:JT, Jensen:2023yxy, AliAhmad:2023etg, Kudler-Flam:2023qfl}, and explains its relevance to the current setting.
	Section \ref{sec:operational} gives a derivation of the main result.
	Section \ref{sec:discussion} discusses extensions of the main result to a more general class of states, and connections to the smooth entropy formalism developed in \cite{renner2004smooth, renner2008security}.
	
	\section{No preferred algebraic entropy for a fluctuating mass}
	\label{sec:no-canonical}
	
	Consider an asymptotically flat Schwarzschild spacetime of mass $M_0$.
	In a full theory of quantum gravity, this spacetime will be described by a state in a Hilbert space $\H_{\text{QG}}$.
	In the semiclassical limit, one can consider the Hilbert space $\H_{\text{fluc}}$ of perturbative fluctuations about the Schwarzschild spacetime.
	This Hilbert space is understood to embed within the full Hilbert space of quantum gravity, $\H_{\text{fluc}} \subseteq \H_{\text{QG}},$ in the sense of quantum error correcting codes \cite{Almheiri:QEC}.
	
	Treating $\H_{\text{fluc}}$ at the quantum level is difficult, but some progress can be made by restricting our attention to the subspace $\H_{\text{mass}}$ of fluctuations to Schwarzschild black holes of mass close to $M_0$.
	Because the Schwarzschild solution has two exteriors, there are two masses that could be described at the quantum level; we will take $\H_{\text{mass}}$ to describe the fluctuations of the mass of the right exterior, and we will treat it as a Hilbert space of square-integrable wavefunctions, $\H_{\text{mass}} \cong L^2(\mathbb{R}).$
	The ``position variable'' on this space is $\delta \hat{M},$ which is the operator that encodes the mass fluctuation.
	It has a conjugate ``momentum'' which we choose to call $\hat{\theta}$ due to the conjugate relationship between black hole mass and boost angle for eternal black holes \cite{Banados:1993qp}.
	
	The algebra of bounded operators on $\H_{\text{mass}}$ is denoted $\B(\H_{\text{mass}}).$
	It is an example of a von Neumann algebra.\footnote{See \cite{Sorce:types} for a thorough exposition of the physical properties of such algebras.}
	The algebra $\B(\H_{\text{mass}})$ is generated by all bounded functions of $\delta \hat{M}$ and $\hat{\theta}.$
	It has a natural subalgebra, $\A_{\text{mass}},$ which is the algebra consisting of all bounded functions of $\delta \hat{M}.$
	This is an abelian von Neumann algebra --- see e.g. \cite[theorem 4.71]{douglas2012banach}.
	
	Let $K$ be a compact subset of the real line.
	The indicator function that is equal to one in $K$ and zero elsewhere is denoted $\chi_K.$
	This is a bounded function, so the operator $\chi_K(\delta \hat{M})$ is in $\A_{\text{mass}},$ and is simply the projection operator onto the $K$-eigenspaces of the Hermitian operator $\delta \hat{M}.$
	We will write $\chi_K(\delta \hat{M}) = \Pi_K.$
	Naively, we might hope to be able to say that at the quantum level, the number of black hole microstates contained in the mass window $K$ is
	\begin{equation} \label{eq:mass-trace}
		\text{dim}(K)
		= \tr(\Pi_K).
	\end{equation}
	But the operator $\Pi_K$ has infinite trace, as is clearly seen in the equation
	\begin{equation} \label{eq:infinite-microstates}
		\tr(\Pi_K)
		= \int dx\, \langle x | \Pi_K | x \rangle = \int_K dx\, \delta(0).
	\end{equation}
	
	To get around a formal infinity in the counting of states, it is necessary to renormalize.
	A method for accomplishing this using the mathematics of von Neumann algebras was proposed in \cite{Witten:crossed-product} and elaborated in \cite{Sorce:types}.
	The idea is to define an abstract \textit{trace} $\tau$, which maps from an ideal $\A_{\text{mass},\tau} \subseteq \A_{\text{mass}}$ of ``trace-class operators'' into the complex numbers $\comps.$
	The trace $\tau$ is defined to be linear and cyclic.
	In order to be a good candidate for a physical renormalized trace, it must also be ``faithful, normal, and semifinite'' --- these conditions are explained in \cite[section 6]{Sorce:types}.
	Given a faithful, normal, and semifinite trace $\tau$ for $\A_{\text{mass}},$ it is possible to define density matrices as the positive operators $\rho \in \A_{\text{mass}}$ that have finite trace, $\tau(\rho) < \infty.$
	The entropy of a density matrix is defined to be
	\begin{equation}
		S_\tau(\rho) = - \tau \left( \frac{\rho}{\tau(\rho)} \log \frac{\rho}{\tau(\rho)} \right).
	\end{equation}
	
	It is easy to find a faithful, normal, and semifinite trace for $\A_{\text{mass}}.$
	A simple one is provided by integration with respect to the Lebesgue measure:
	\begin{equation}
		\tau_{0}(f(\delta \hat{M}))
		= \int dx\, f(x).
	\end{equation}
	The corresponding trace-class ideal $\A_{\text{mass}, \tau_0}$ is just the space of Lebesgue-integrable functions $L^1(\mathbb{R}).$
	The trouble is that there are \textit{too many} faithful, normal and semifinite traces on $\A_{\text{mass}},$ and none is any better than the rest for defining entropy!
	In fact, for any positive function $\mu(x),$ the functional
	\begin{equation}
		\tau_{\mu}(f(\delta M))
		= \int dx\, \mu(x) f(x)
	\end{equation}
	is a faithful, normal, semifinite trace \cite[page 322]{TakesakiI}.\footnote{A classic theorem \cite[theorem V.2.31]{TakesakiI} implies that every faithful, normal, semifinite trace on $\A_{\text{mass}}$ is realized in this way.}
	The formal statement is that because $\A_{\text{mass}}$ is an abelian algebra, it is type I (see \cite[proposition V.1.23]{TakesakiI}), but because it has nontrivial center, it is not a factor (see \cite[section 3]{Sorce:types}), leading to a large ambiguity in the appropriate definition of a trace.
	
	The ambiguity in the trace $\tau_{\mu}$ makes it hard to come up with anything we might canonically define as the entropy of a density matrix for mass fluctuations.
	Even if we were to insist on a measure we particularly like, such as the Lebesgue measure, our resulting formula for entropy would have no physical interpretation.
	To remedy this, we will briefly review in the next section how coupling the black hole mass to quantum fields yields a preferred family of renormalized traces and show that this family of traces, when pulled back to $\A_{\text{mass}}$, corresponds to the family of traces $\tau_{\mu}$ where $\mu(x)$ is proportional to $e^x$.
	The advantage of this is that we will then be able to show in section \ref{sec:operational} that entropy differences of microcanonical mass wavefunctions actually do have an interpretation in terms of relative state-counting in the semiclassical algebra.
	
	\section{Gravity and the crossed product}
	\label{sec:crossed-product}
	
	Consider now a Hilbert space $\H_{\text{QFT}}$ of quantum fields on the background of an eternal Schwarzschild black hole, such that this Hilbert space includes the Hartle-Hawking state $|\Psi\rangle$ which is in equilibrium with respect to Schwarzschild time evolution \cite{hartle1976path}.\footnote{This setting is particularly natural for Schwarzschild-AdS black holes.
		For asymptotically flat black holes it may be more natural instead to consider the Hilbert space containing the Unruh state.
		All of the calculations in this section were carried out for the Unruh background in \cite{Kudler-Flam:2023qfl}.}
	Consider the von Neumann algebra $\A_{\text{QFT}}$ generated by fields in the right exterior of the black hole.
	Naively, the Hilbert space of quantum fields coupled to mass fluctuations is
	\begin{equation}
		\H
		= \H_{\text{QFT}} \otimes \H_{\text{mass}},
	\end{equation}
	and the corresponding algebra of operators is
	\begin{equation} \label{eq:unconstrained-algebra}
		\A
		= \A_{\text{QFT}} \otimes \B(\H_{\text{mass}}).
	\end{equation}
	In physical states, however, there is a relationship between the fluctuation of the mass and the fluctuation of the stress-energy tensor of the quantum fields.
	If $\xi^a$ is the Schwarzschild time vector field that is future-directed in the right exterior, then the physical constraint is enforced by the operator equation
	\begin{equation}
		M_{\text{right exterior}} - M_{\text{left exterior}}
		= \int_{\Sigma} T_{ab} \xi^a d\Sigma^b,
	\end{equation}
	where $\Sigma$ is any Cauchy slice of the full Schwarzschild spacetime, and $T_{ab}$ is an appropriately renormalized stress-energy tensor for quantum fields. 
	The physical operators in the theory are the ones that commute with the constraint
	\begin{equation}
		C = M_{\text{right exterior}} - M_{\text{left exterior}} - \int_{\Sigma} T_{ab} \xi^a d \Sigma^b.
	\end{equation} 
	Equation \eqref{eq:unconstrained-algebra} must therefore be modified to obtain a physical algebra of operators by restricting to the subalgebra
	\begin{equation}
		\hat{\A}
		= \left\{\mathrm{a} \in \A\, | \, [\delta \hat{M} - \int_{\Sigma} T_{ab} \xi^a d\Sigma^b, \mathrm{a}] = 0 \right\}.
	\end{equation}
	The integral $\int_{\Sigma} T_{ab} \xi^a d\Sigma^b$ is proportional to the modular Hamiltonian of the Hartle-Hawking state $|\Psi\rangle$ \cite{hartle1976path}.
	The algebra $\hat{\A}$ is called a crossed product.
	
	Crucially, it was shown in \cite{takesaki1973duality, Witten:crossed-product} (see also \cite[appendix B]{Jensen:2023yxy}) that the crossed product is a type II von Neumann factor.
	As explained in \cite[section 6]{Sorce:types}, a type II von Neumann factor has a single, $\reals^+$-parametrized family of renormalized traces, all of which are related by a single constant ambiguity in scaling.
	One member of this family is \cite[appendix B]{Jensen:2023yxy}
	\begin{equation}
		\tau(\hat{\mathrm{a}})
		= 2 \pi \langle \Psi| \langle 0|_{\theta} e^{\delta \hat{M}/2} \hat{\mathrm{a}} e^{\delta \hat{M}/2} |0\rangle_{\theta} |\Psi\rangle,
	\end{equation}
	where $|\Psi\rangle$ is the Hartle-Hawking state and $|0\rangle_{\theta}$ is the zero eigenstate of the ``momentum'' operator $\hat{\theta}.$
	All other faithful, normal, semifinite traces are of the form $c \tau$ for $c > 0.$
	
	The mass fluctuation algebra $\A_{\text{mass}}$ is a subalgebra of the gauge-invariant algebra $\hat{\A}.$
	The canonical family of traces on $\hat{\A},$ when pulled back to $\A_{\text{mass}},$ give
	\begin{equation}
		c \tau(f(\delta \hat{M}))
		= c \int dx\, e^{x} f(x).
	\end{equation}
	They are therefore all proportional to rescalings of the Lebesgue measure by the smooth function $e^x,$ as was claimed in section \ref{sec:no-canonical}.
	Given any wavefunction $g(x)$ for the mass fluctuation, one constructs a corresponding state
	\begin{equation}\label{eq:gen_mass_state}
		|g\rangle_{\text{mass}}
		= \int dx\, g(x) |x\rangle_{\text{mass}},
	\end{equation}
	and defines the corresponding density matrix $\rho_g$ with respect to the trace $\tau$ as the operator $\rho_g$ in $\A_{\text{mass}}$ satisfying
	\begin{equation}
		\tau(\rho_g f(\delta \hat{M}))
		= \langle g| f(\delta \hat{M}) | g \rangle_{\text{mass}}
	\end{equation}
	for every operator $f(\delta \hat{M}) \in \A_{\text{mass}}.$
	A straightforward calculation shows that $\rho_g$ is unique and is given by
	\begin{equation} \label{eq:density-matrix}
		\rho_g
		= |g(\delta \hat{M})|^2 e^{- \delta \hat{M}}.
	\end{equation}
	For any such density matrix one can formally define and compute the entropy
	\begin{eqnarray}
		- \tau(\rho_g \log \rho_g)
		& = & \int dx\, x |g(x)|^2 - \int dx\, |g(x)|^2 \log|g(x)|^2 \nonumber \\
		& = & \langle \delta \hat{M} \rangle_{g} + S_g.
	\end{eqnarray}
	Because the normalization of $\tau$ is ambiguous up to rescaling, entropy is only defined up to a constant, and the mathematically meaningful quantity that can be computed is
	\begin{eqnarray} 
		&& - \tau(\rho_{g_2} \log \rho_{g_2})
		+ \tau(\rho_{g_1} \log \rho_{g_1}) \nonumber \\
		&& \quad\qquad = (\langle \delta \hat{M} \rangle_{g_2} -  \langle \delta \hat{M} \rangle_{g_1}) + (S_{g_2} - S_{g_1}). \label{eq:entropy-difference}
	\end{eqnarray}
	This is more or less what was done in \cite{Chandrasekaran:dS, Chandrasekaran:large-N, Penington:JT, Jensen:2023yxy, AliAhmad:2023etg, Kudler-Flam:2023qfl}, except that there entropy differences were computed for states of the form $|\Phi\rangle_{\text{QFT}} \otimes |g\rangle_{\text{mass}}$ in the full matter-mass Hilbert space.
	The case being considered here is the special case where the quantum field theory state is fixed to the Hartle-Hawking state $|\Psi\rangle,$ and only the mass wavefunction changes.
	
	\section{Operational entropy differences for microcanonical states}
	\label{sec:operational}
	
	Let $K$ be a compact subset of the real line, and consider the microcanonical wavefunction
	\begin{equation}\label{eq:micro_mass_states}
		|K\rangle_{\text{mass}}
		= \frac{\int_K dx\, e^{x/2} |x\rangle_{\text{mass}}}{\sqrt{\int_K dx\, e^{x}}}.
	\end{equation}
	The weighting $e^{x/2}$ is chosen so that the associated density matrix in the type II algebra, given by equation \eqref{eq:density-matrix}, will be proportional to a projection operator --- this is what it should mean for a state to be microcanonical.
	For two such sets $K_1$ and $K_2,$ equation \eqref{eq:entropy-difference} gives the type II entropy difference as
	\begin{equation}
		S_{\text{II}}(K_2)
		- S_{\text{II}}(K_1)
		= \log \frac{\int_{K_2} dx\, e^{x}}{\int_{K_1} dx\, e^{x}}.
	\end{equation}
	Because of our choice of microcanonical wavefunction, this is equal to the logarithm of a ratio of traces of projection operators in $\A_{\text{mass}}$:
	\begin{equation}
		S_{\text{II}}(K_2)
		- S_{\text{II}}(K_1)
		= \log \frac{\tau(\Pi_{K_2})}{\tau(\Pi_{K_1})}.
	\end{equation}
	This fact will allow us to give the entropy difference a relative state-counting interpretation.
	Without loss of generality, let us assume $\tau(\Pi_{K_2}) \geq \tau(\Pi_{K_1}).$
	Let us write the ratio as an integer $n$ plus a remainder term that is less than one:
	\begin{equation}
		\frac{\tau(\Pi_{K_2})}{\tau(\Pi_{K_1})}
		= n + r.
	\end{equation}
	In any type II factor, it is a theorem --- see e.g. \cite[proposition 1.3.5]{sunder2012invitation} or \cite[page 32]{Sorce:types}--- that the support of $\Pi_{K_2}$ can be decomposed into $n$ orthogonal Hilbert spaces that are ``equivalent'' to the image of $\Pi_{K_1},$ plus a piece corresponding to the remainder term.
	Concretely, this means that there exist projectors $\tilde{\Pi}_{1}, \dots \tilde{\Pi}_{n}, \tilde{\Pi}_{r} \in \hat{\A}$ with
	\begin{equation}
		\Pi_{K_2}
		= \tilde{\Pi}_1 + \dots + \tilde{\Pi}_{n} + \tilde{\Pi}_{r},
	\end{equation}
	with each of the first $n$ projectors equivalent to $\Pi_{K_1}$ in the sense that there exists a partial isometry $V_j \in \hat{\A}$ with
	\begin{equation}
		V_j \Pi_{K_1} V_j^{\dagger}
		= \tilde{\Pi}_{j}, \quad V_j^{\dagger} \tilde{\Pi}_{j} V_j = \Pi_{K_1}.
	\end{equation}
	In the special case $r=0$ this is the whole story; for $0 < r < 1,$ the remainder piece $\tilde{\Pi}_{r}$ can be embedded within an $(n+1)$-st projector equivalent to $\Pi_{K_1}$:
	\begin{equation}
		\tilde{\Pi}_{r} \subsetneq \tilde{\Pi}_{n+1} = V_{n+1} \Pi_{K_1} V_{n+1}^{\dagger}, \quad V_{n+1}^{\dagger} \tilde{\Pi}_{n+1} V_{n+1} = \Pi_{K_1}.
	\end{equation}
	As explained in \cite[section 5]{Sorce:types}, all nontrivial projectors in a type II factor have infinite rank, so the existence of a partial isometry relating any two projectors is not special; what is special is that the partial isometry relating the projectors lives within the type II factor.
	In the context of black hole physics, this means that the microcanonical window of black hole mass on which $\Pi_{K_2}$ is supported can be broken up into $n$ pieces that are isomorphic to the support of $\Pi_{K_1}$ \textit{with the isomorphism implemented by a gauge-invariant operator}, plus a remainder term.
	There is also a theorem (\cite[proposition 1.3.5]{sunder2012invitation}) stating that $\Pi_{K_2}$ cannot be decomposed into any different number $m$ of projectors equivalent to $\Pi_{K_1}$ --- the integer $n,$ plus the remainder term, is unambiguously determined by the renormalized trace.
	It is important to note that because the operators $V_j$ act on the full semiclassical Hilbert space $\H_{\text{QFT}} \otimes \H_{\text{mass}}$, the projections $\tilde{\Pi}_j$ can act nontrivially on the quantum matter degrees of freedom, even though $\Pi_{K_2}$ and $\Pi_{K_1}$ act only on $\H_{\text{mass}}$.
	
	This result can be put in a more suggestive form by introducing an auxiliary finite-dimensional Hilbert space $\comps^{N},$ with $N=n$ for $r=0,$ and $N=n+1$ for $r > 0.$
	In the case $r=0,$ the operator
	\begin{equation}
		U = \sum_{j=1}^{N} V_j \otimes \langle j |
	\end{equation}
	provides a unitary equivalence between $\H_{\text{QFT}} \otimes \H_{K_1} \otimes \comps^N$ and $\H_{\text{QFT}} \otimes \H_{K_2},$ with $\H_{K} \equiv \Pi_K \H_{\text{mass}}.$
	In the case $r>0,$ $U$ provides a unitary equivalence between $\H_{\text{QFT}} \otimes \H_{K_1} \otimes \comps^N$ and a slightly enlarged Hilbert space containing $\H_{\text{QFT}} \otimes \H_{K_2}.$
	In either case, the microcanonical Hilbert space $\H_{\text{QFT}} \otimes \H_{K_2}$ can be embedded in the finitely-augmented microcanonical Hilbert space $\H_{\text{QFT}} \otimes \H_{K_1} \otimes \comps^N$ using only operators in the type II gauge-invariant algebra, and $N$ is the smallest integer for which such an embedding is possible.
	This is the sense in which the type II entropy difference for microcanonical mass wavefunctions represents a relative counting of states.
	
	\section{More general states}
	\label{sec:discussion}
	
	So far, we have explained a relative state-counting interpretation for type II entropy differences between states of the form $|K\rangle_{\text{mass}}$ (cf. equation \eqref{eq:micro_mass_states}).
	What about more general states?
	For a general (i.e. non-microcanonical) state in a finite-dimensional Hilbert space, the size of the support is not quantified by the von Neumann entropy, but by the \emph{max-entropy},
	\begin{equation}
		S_\mathrm{max}(\rho) = \log \rank \rho.
	\end{equation}
	The natural generalization of this quantity to the type II setting is 
	\begin{equation}
		S_\mathrm{max}(\rho) = \log \tau(\rho^0).
	\end{equation}
	The states \eqref{eq:micro_mass_states} were chosen specifically because their von Neumann entropy differences and max entropy differences are equal.
	For a general mass wavefunction $|g\rangle_{\text{mass}},$ it is easy to see from equation \eqref{eq:density-matrix} that the type II max-entropy is $\log \tau(\Pi_{\text{supp}(g)}).$
	For compactly supported mass wavefunctions this is finite, and the type II max-entropy difference has the same operational interpretation as the von Neumann entropy had for the microcanonical states \eqref{eq:micro_mass_states}.
	
	For wavefunctions that are not compactly supported, the max-entropy can be infinite, and the interpretation of max-entropy difference in terms of relative state counting is less useful.
	There is however still a useful quantity for understanding \textit{approximate} relative state counting in the form of the \textit{smooth max-entropy}:\footnote{In the finite-dimensional setting some authors use different notions of distance in the infimum, but these do not cause an essential change in the physics.}
	\begin{equation} \label{eq:smoothing}
		S^\varepsilon_\mathrm{max}(\rho) = \inf_{\tau(|\hat{\rho} - \rho|) \leq \epsilon} S_\mathrm{max}(\hat{\rho}).
	\end{equation}
	H\"{o}lder's inequality holds for any renormalized trace \cite{nelson1974notes}, and implies that for any operator $O$ in the type II algebra with operator norm $\lVert O \rVert_{\infty}$, the difference in expectation values between $\hat{\rho}$ and $\rho$ satisfies
	\begin{equation}
		|\tau(\rho O) - \tr(\hat{\rho} O)| \leq \tau(|\hat{\rho} - \rho|)\, \lVert O \rVert_{\infty}.
	\end{equation}
	Operators $\hat{\rho}$ appearing in the infimum in equation \eqref{eq:smoothing} therefore approximate $\rho$ in an operational sense, and the smooth max-entropy may be thought of as measuring the max-entropy of the ``important'' part of $\rho.$
	
	It is easy to see from equation \eqref{eq:density-matrix} that for a generic mass wavefunction $|g\rangle_{\text{mass}},$ the infimum in equation \eqref{eq:smoothing} will be attained by a truncation of $\rho_g$ onto a subspace of the mass Hilbert space with finite renormalized trace.
	Given two such density matrices $\rho_{g_1}$ and $\rho_{g_2}$ with (assumed positive) smooth max-entropy difference
	\begin{equation}
		S^\varepsilon_\mathrm{max}(\rho_{g_2}) - S^\varepsilon_\mathrm{max}(\rho_{g_1}) = \Delta S_\mathrm{max}^\varepsilon,
	\end{equation}
	and letting $\hat{\Pi}_{g_1}$ and $\hat{\Pi}_{g_2}$ be the projectors onto the truncated subspaces, our previous considerations for microcanonical states imply that there is gauge-invariant unitary embedding of $\mathcal{H}_\mathrm{QFT} \otimes \hat{\Pi}_{g_2} \mathcal{H}$ into $\mathcal{H}_\mathrm{QFT} \otimes \hat{\Pi}_{g_1} \mathcal{H}_{\text{mass}} \otimes \mathbb{C}^n$ if and only if $n \ge e^{\Delta S_\mathrm{max}^\varepsilon}$.
	
	In the full semiclassical Hilbert space $\H_{\text{QFT}} \otimes \H_{\text{mass}},$ the states we have described are ones for which the quantum field theory is fixed to be the Hartle-Hawking state $|\Psi\rangle$, and only the mass wavefunction changes.
	Entropy differences in the more general setting where the state of the quantum fields can change are governed by the formulas of \cite{Chandrasekaran:dS, Chandrasekaran:large-N, Penington:JT, Jensen:2023yxy, AliAhmad:2023etg, Kudler-Flam:2023qfl}.
	It would be interesting to understand the physics of smooth max-entropy differences for such states, and to relate this to previous work on smooth entropies in quantum gravity \cite{Bao:smooth, Akers:smooth1, Akers:smooth2, Akers:smooth3}.
	
	\section*{Acknowledgments}
	This work benefited from conversations with Netta Engelhardt and Jonah Kudler-Flam.
	CA is supported by National Science Foundation under the grant number PHY-2207584, the Sivian Fund, and the Corning Glass Works Foundation Fellowship.
	JS is supported by the AFOSR under award number FA9550-19-1-0360, by the DOE Early Career Award number DE-SC0021886, and by the Heising-Simons Foundation.

	\bibliographystyle{JHEP}
	\bibliography{bibliography}

\end{document}